\documentclass[sigconf, screen]{acmart}

\acmConference[PREPRINT]{PREPRINT}{June 2025}{}
\usepackage[american]{babel}

\usepackage{xspace}
\usepackage{graphicx}
\usepackage{microtype}
\usepackage{hyperref}
\usepackage{csquotes}
\usepackage{amsmath, amsfonts}
\usepackage{physics}
\usepackage{siunitx}
\usepackage{cleveref}
\usepackage{qcircuit}
\usepackage{subcaption}
\usepackage{placeins}

\begin{document}

\title{Quantum Computer Fingerprinting using Error Syndromes}
% \title{Security through Recycling: Fingerprinting Quantum Computers with Error Syndromes (TODO)}
% \title{Recycling Error Syndromes to Fingerprint Quantum Computers}

% TODO: De-anonymize

\author{Vincent Mutolo}
\email{vm2724@columbia.edu}
\orcid{0009-0005-8717-842X}
\affiliation{%
  \institution{Columbia University}
  \city{New York}
  \state{NY}
  \country{USA}
}

\author{Devon Campbell}
\email{dec2180@columbia.edu}
\orcid{}
\affiliation{%
  \institution{Columbia University}
  \city{New York}
  \state{NY}
  \country{USA}
}

\author{Quinn Manning}
\email{jqm2108@columbia.edu}
\orcid{0009-0005-5373-3843}
\affiliation{%
  \institution{Columbia University}
  \city{New York}
  \state{NY}
  \country{USA}
}

\author{Henri Witold Dubourg}
\email{henri.dubourg@columbia.edu}
\orcid{0009-0008-9336-8774}
\affiliation{%
  \institution{Columbia University}
  \city{New York}
  \state{NY}
  \country{USA}
}

\author{Ruibin Lyu}
\email{rl3501@columbia.edu}
\orcid{}
\affiliation{%
  \institution{Columbia University}
  \city{New York}
  \state{NY}
  \country{USA}
}

\author{Simha Sethumadhavan}
\email{simha@columbia.edu}
\orcid{}
\affiliation{%
  \institution{Columbia University}
  \city{New York}
  \state{NY}
  \country{USA}
}

\author{Dan Rubenstein}
\email{danr@cs.columbia.edu}
\orcid{}
\affiliation{%
  \institution{Columbia University}
  \city{New York}
  \state{NY}
  \country{USA}
}

\author{Salvatore Stolfo}
\email{sal@cs.columbia.edu}
\orcid{0000-0003-1611-0100}
\affiliation{%
  \institution{Columbia University}
  \city{New York}
  \state{NY}
  \country{USA}
}

\begin{abstract}
As quantum computing matures and moves toward broader accessibility through cloud-based platforms, ensuring the authenticity and integrity of quantum computations becomes an urgent concern. In this work, we propose a strategy to leverage the byproducts of quantum error correction (QEC) to verify hardware identity and authenticate quantum computations for ``free'', without introducing any additional quantum computations or measurements. By treating syndrome measurements as a source of metadata, we embed verification seamlessly into standard QEC protocols and eliminate the need for separate challenge-response pairs. We validate our approach using multiple error-correcting codes, quantum states, and circuit compilation strategies on several generations of IBM quantum computers. Our classifiers achieve 99\% accuracy with only 500 shots in distinguishing among five backends. Overall, we re-purpose the intrinsic overhead of error correction to be a mechanism for securing quantum computation.
\end{abstract}

  
\maketitle

\section{Introduction}

Quantum computing is a nascent and rapidly evolving technology with the potential to revolutionize computation \cite{ladd2010quantum,montanaro2016quantumalgos,mikenike,feynman1982sim,harrow2017supremacy}. As quantum technologies advance toward practical implementation, ensuring the authenticity and integrity of quantum computations becomes increasingly critical. Quantum hardware exhibits unique characteristics, such as superposition and entanglement, that not only empower its computational capabilities but also introduce complex challenges in verification and identity authentication. 

Traditional methods for verifying computations and authenticating hardware identities may be insufficient in the quantum realm due to the relatively high cost of running quantum circuits and their inherent coupling with their changing environment. Quantum devices are susceptible to physical drift over time caused by environmental factors, hardware imperfections, and quantum decoherence \cite{proctor2020drift}. This drift affects the performance and the characteristic fingerprints of the quantum hardware, complicating long-term verification.

In this paper, we propose a framework that intrinsically ties the verification of quantum computations to the computations themselves. We present a method to authenticate quantum hardware by leveraging the byproducts of quantum error correction (QEC). This approach offers three key advantages: 
\begin{enumerate}
    \item It requires no additional quantum circuitry or measurement beyond the existing quantum computation infrastructure, as QEC is already necessary for meaningful quantum operations \cite{campbell2017faulttolerant}.
    \item The information used for verification is ``free'' in the sense that it was already measured and is available in classical form, as opposed to being maintained in a quantum state that would require additional measurement.
    \item The verification fingerprint is computed based on a portion of the business circuit, fundamentally coupling the fingerprint and the circuit. This means it is more difficult to present a legitimate verification for a circuit run dishonestly by only running the verification honestly.
\end{enumerate}

Quantum computers have significantly higher error rates than classical computers. This fundamental difference arises due to the fragility of quantum states, which are affected by noise and environmental disturbances throughout their lifecycle from initialization to transformation to readout \cite{gottesman2010errorcorrection,mikenike}.

Quantum error correction encodes logical qubits using entangled states of multiple physical qubits and must frequently perform quantum computation upon these logical qubit structures to produce {\it error syndromes}: classical bit strings that diagnose and correct the types and locations of errors without collapsing the quantum state of the logical qubits. As we will demonstrate, these syndrome measurements offer an inexpensive and powerful method of authentication. Each distinct quantum computer produces a distinct statistical mix of syndromes, which we use to build a fingerprinting system.

In classical computing, physical unclonable functions (PUFs) are used to verify the identity of physical devices (often through a proxy, such as verifying interaction with a physical circuit) \cite{gassend2002silicon}. Early PUFs were based on lasers passing through an optical medium that would cause complex and unpredictable scattering; the resulting pattern could be used as a fingerprint \cite{pappu2002puf}. Now, classical PUF designs \cite{holcomb2009srampuf,sakhare2021ringpuf,guajardo2007ip,edward2007ring,gassend2002silicon} are often tied intrinsically to the hardware we care about verifying, but tend to require out-of-band verification protocols --- that is, challenge-response pairs (CRPs) are exchanged that are irrelevant to any useful computation. In the context of quantum computing, this is both expensive, requiring extra circuitry on a platform that costs \$96 per minute \cite{ibmPricingQuantum}, and means the verification is decoupled from the business circuit, which enables a dishonest provider that can detect which circuits are part of the PUF challenge to lie to users except when the request is for verification.  Our goal is to not only prove an entity has \emph{access} to a device, but also to prove that it \emph{used} this device for a specific computation.

Several quantum-specific PUFs (QPUFs) have been proposed; some require the client and provider to share a quantum communication channel \cite{skoric2009qreadout,goorden2014qreadout,blank2020quantum}, while others require running circuits dedicated to the verification protocol \cite{mi2021short, phalak2021puf, smith2023fast, bathalapalli2023qpuf}, which is both expensive and decouples the proof from the computation we want to verify.

Treating syndrome measurements as a source of metadata about the physical systems running our computation, our approach employs machine learning classifiers (neural networks \cite{sklearn,pytorch}) to find patterns and correlations characteristic of specific quantum computers. This fingerprinting method is difficult to detect; we show that the outputs of circuits with no intentional fingerprinting routines carry enough information in syndrome measurements to characterize backends (the quantum computers companies use to provide quantum cloud services). Our proposed protocol does not require participation from the prover, which is an essential property of a system that addresses routing attacks, which we define as the prover detecting whether a circuit is used for authentication, and then running that circuit honestly, while possibly acting dishonestly for other inputs.

To validate our framework, we conduct a series of experiments on simple circuits varying the error correcting code, logical-to-physical qubit mapping (hereafter just called ``mapping''), starting state, and compilation strategy used for each circuit; we demonstrate that our method is robust over variations in these parameters. We perform these experiments on several generations of IBM quantum computers. Our results demonstrate that syndrome measurements contain information that enables classifiers to accurately identify the backend in use with significantly higher accuracy than random guessing. Our classifiers achieve over 99\% accuracy in distinguishing between five IBM backends with fewer than 500 shots (measured samples of a quantum circuit). 

% (TODO: Do we even still want this?) Moreover, we address the challenge of quantum hardware drift by modeling how these changes affect the verification process. Recognizing that the physical properties of quantum devices evolve over time, we demonstrate that models based on initial measurements of hardware characteristics become less reliable, leading to potential verification failures. To mitigate this, we introduce strategies for retraining the verification model, allowing it to adapt to the hardware's evolution.

In summary, this work introduces a framework that leverages native quantum error correction protocols to provide device identification and authentication without incurring the cost or vulnerability of separate “out-of-band” verification. Specifically, this paper contributes the following:

\begin{enumerate}
    \item We demonstrate that error syndrome measurements carry information about quantum hardware, showcasing the feasibility of using these measurements for verification fingerprints. (\cref{sec:classiyfing-backends}).

    \item We study the resilience of our approach to changes in error correcting codes (\cref{sec:vary-codes}), varied qubit mappings (\cref{sec:vary-mappings}), varied starting states and layout methods (\cref{sec:vary-states-layout}), and drift over time (\cref{sec:stability}).

    \item We propose a protocol for using error syndromes to address the two real-world threat models defined in \cref{sec:threat-model}. This is meant to be a proof of concept rather than a rigorous protocol.

    \item We discuss the power of two different physical models of the underlying hardware to predict our results, specifically whether backends can be distinguished by error syndromes. (\cref{sec:causal})
\end{enumerate}

\paragraph{Paper Organization} \Cref{sec:background} provides background on quantum concepts the reader may find helpful for understanding our results; \cref{sec:threat-model} introduces the real-world threat models that motivate our authentication mechanism; \cref{sec:method} proposes a method to fingerprint quantum computers under those threat models; \cref{sec:results} discusses our experiments and results that demonstrate the performance of our method;  \cref{sec:limitations} discusses the security offered by our construction and limitations of our current results; \cref{sec:related-work} provides related work in both the classical and quantum domains and discusses where our contributions fit in;\cref{sec:conclusion} concludes.

Quantum computation intrinsically requires expensive error correction routines \cite{campbell2017faulttolerant}. Instead of adding further expensive computation to verify hardware identity, our method provides verification for ``free'' by reusing error syndromes as fingerprints for the underlying hardware.

% takeaway blurb

\section{Background} \label{sec:background}

The qubits that make up quantum computers are extremely sensitive; in fact, there is a large body of research under the term ``quantum sensing'' that studies how various physical realizations of qubits may be considered effective sensors for external systems \cite{degen2017sensing}. However, in the context of quantum computing, their sensitivity is a troublesome property that disrupts computation; this leads to expensive engineering to shield the quantum computer from external influence and the development of error-correction mechanisms to make the system robust to inevitable leaks in physical isolation and miscalibration of control devices.

Quantum error correcting codes (QECs) are therefore crucial for the development of reliable quantum computing \cite{campbell2017faulttolerant}. QEC implementations often rely on additional ancilla qubits that we will refer to as {\em syndrome} qubits, included specifically into the circuit to characterize possible errors. The QEC performs computations to set these syndrome qubits and performs syndrome measurements on these syndrome qubits taken throughout a computation. These measurements on the syndrome qubits do not reveal any information about the state of the non-syndrome, data-carrying qubits (which would otherwise generally destroy the quantum computation), but instead describe the errors that have occurred such that these errors can be corrected by applying appropriate operations upon the data qubits that remain in their quantum state.

\begin{figure}
    \centering
\[
    \Qcircuit @C=1em @R=1em {
    & \lstick{\ket{\psi}} & \qw & \ctrl{1} & \ctrl{2} & \multigate{2}{\varepsilon} & \ctrl{3} & \qw & \qw & \qw & \qw \\
    & \lstick{\ket{0}}    & \qw & \targ    & \qw      & \ghost{\varepsilon}        & \qw & \ctrl{2} & \ctrl{3} & \qw & \qw \\
    & \lstick{\ket{0}}    & \qw & \qw      & \targ    & \ghost{\varepsilon}        & \qw & \qw & \qw & \ctrl{2} & \qw \\
    & \lstick{\ket{0}}    & \qw & \qw & \qw & \qw & \targ & \targ & \qw & \qw & \meter & \cw & \rstick{b_0} \\
    & \lstick{\ket{0}}    & \qw & \qw & \qw & \qw & \qw & \qw & \targ & \targ & \meter & \cw & \rstick{b_1} \\
    }
\]
    \caption{The quantum bit flip repetition code. The top three qubits are considered ``data'' qubits in that together they carry the information about the original state $\ket{\psi}$. The bottom two qubits are considered ``syndrome'' qubits in that their purpose is to diagnose the errors in the error channel represented here by the gate $\varepsilon$. \cite{mikenike}}
    % extra caption: The first syndrome qubit is essentially a parity measurement of the first two data qubits. The second syndrome is essentially a parity measurement of the bottom two data qubits. These two parity measurements together are enough to diagnose and correct a single bit flip error. Note that $b_0,b_1$ above are \emph{classical} bits, and therefore comparatively very cheap to work with. They can be freely copied out of the system without disrupting any further quantum computation. 
    \label{fig:q-rep-code}
\end{figure}
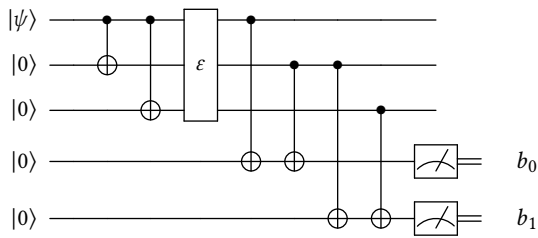

% TODO: Get rid of this example if we need space.
For example, one of the simplest QECs is the bit-flip repetition code, which works by measuring the parity between bits. If the parity comes back 0 for all pairs of bits, we assume there has been no bit flip (or two bit flips, which is considered out-of-scope for this code to handle). If the parity comes back 1 for some pairs, we can deduce which qubit flipped based on which pairs have a parity of 1. \cite{mikenike} The quantum bit-flip repetition code is presented in \cref{fig:q-rep-code}. Information about \emph{errors} will be measured and leave the system while the quantum computation progresses undisturbed. These errors can either be corrected during computation by adding a bit flip gate conditioned on the parity measurements, or after the computation has finished by reinterpreting measurement results to account for the error \cite{roffe2019error}.  \Cref{fig:q-rep-code} shows an implementation of this code that may be run as a subcircuit during a quantum computation. The most important takeaway is that the top three data qubits remain unmeasured through the error detection process, and that the bottom two syndrome qubits output classical bit strings representing which data qubit, if any, experienced a bit flip error.

Quantum errors are unfortunately not confined to bit-flip errors. In fact, there are infinite errors that may occur to a qubit; its state space is continuous. Fortunately, through the act of measuring the syndrome qubits, the error on any given data qubit is \emph{collapsed} to be one of four possible states: no error, a bit flip error, a \emph{phase} flip error, or both a bit flip and a phase flip error. That is, a single-qubit error will be collapsed through measurement to the composition of up to two errors: an $X$ error or a $Z$ error \cite{mikenike}.

The repetition code presented in \cref{fig:q-rep-code} only corrects one of the two kinds of errors we may observe. (Which one it corrects is arbitrary.) Error codes also have a certain tolerance for the error of the underlying qubits, and certain codes have a higher tolerance than others; the tolerance of the repetition code is not particularly good. For these and other reasons, more complex and more performant error correcting codes are used in practice, such as the surface code \cite{fowler2012surface}. However, QECs generally share some characteristics that we will assume for the purposes of this paper: Error correcting codes use data qubits and syndrome qubits; through quantum computation, syndrome qubits are given information about the errors that have occurred during computation (often in the form of parity checks). These syndromes are then measured and output as classical bit strings that can be decoded to find and optionally correct the errors that occurred, all without collapsing the delicate quantum state carried through the computation (only the errors themselves are collapsed).

\section{Threat Model} \label{sec:threat-model}

Consider the case of a {\em business} wanting to run a quantum computation. The business in all likelihood will need to rent time on a quantum computer of a {\em provider}. The business's use case will dictate some target error rate for the computation, and the business will select an offering that provides access to a specific quantum computer (which we will call a ``backend'') that meets the target error rate at an acceptable cost. The business wants to ensure that the circuits it submits to the provider \emph{really do} run on the machine it is paying for.

We consider the following threat models representing situations where a user may be deceived regarding what quantum device they believe to be hosting their circuit. In all cases, we assume a client-server model for quantum computation, where users or businesses (clients) submit a quantum circuit to a provider (server); the provider has a choice of some number of backends on which to run the quantum circuit. We assume that no classical computer can efficiently simulate quantum computations, or, similarly, we consider only quantum circuits large enough to be impossible to simulate classically.

\begin{description}
    \item[Overloaded provider] Some providers may \emph{usually} be honest and route submitted circuits to their intended backends, but purposefully choose a less expensive backend to reduce queue sizes on more favorable machines. In this model, we cannot assume the provider will always honestly cooperate in verification. 
    
    \item[Dishonest reseller] Quantum computing providers may empower third-party organizations to sell compute time on behalf of first-party owners of quantum computers. In this scenario, users may trust the first-party providers but not the third-party resellers. And in this case, we may assume the first-party provider will honestly cooperate in verification.

\end{description}

Our work demonstrates the feasibility of fingerprinting backend systems under both of these models, i.e.~with or without cooperation from the first-party hosting provider. The important capabilities that enable this are (1) preventing providers from detecting that a circuit will be used for authentication, and (2) being robust to certain transformations providers can perform on user-submitted circuits, such as modifying the mapping from logical to physical qubits.

\section{Using Syndrome Measurements}
\label{sec:method}

% reference diagram somewhere

\begin{figure*}
    \centering
    \includegraphics[width=1\linewidth]{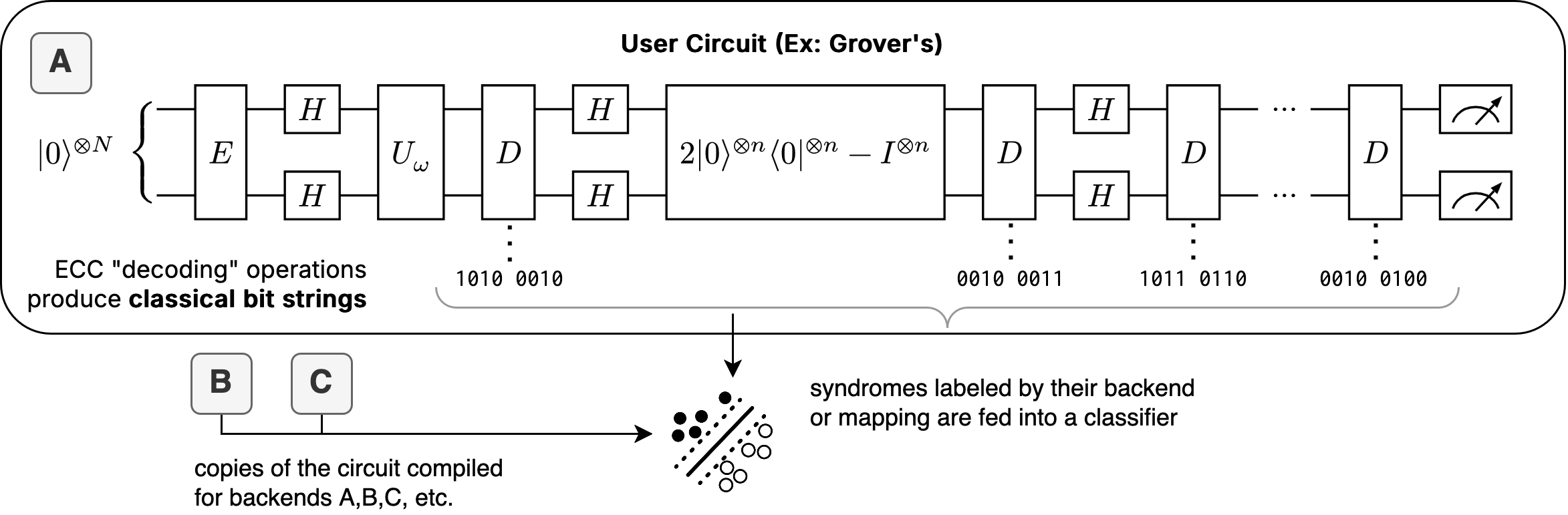}
    \caption{Training begins with selecting a representative circuit. Above, Grover's algorithm \cite{grover1996grover} (with error-correcting encoding and decoding applied, represented respectively by $E$ and $D$ gates) is our example representative circuit. The circuit is transpiled for several different backends, labeled $A$, $B$, and $C$ above. The circuit is run some fixed number of times per backend, say 5000 times. The syndrome measurements, representing the errors that have been observed throughout the circuit's execution, are fed into an SVM classifier.}
    \label{fig:training}
\end{figure*}

Information about intrinsic hardware variations can often be used to create a PUF. One example from the classical world is reading the SRAM ``power-up'' state, as proposed by Holcomb~et~al.~\cite{holcomb2009srampuf}. Whether the bits are initialized to one or zero on power-up is determined by manufacturing process variation. Similarly, since qubits are extremely sensitive states \cite{degen2017sensing}, we can expect this to be especially true for quantum computers.

Here we come to the intuition for our method. We recognize that quantum error correcting codes output (classical) information that describes the hardware errors that occurred on the quantum device. Hardware errors are indicative of process variation (since one typically does not intend for errors to occur), and we can fingerprint on this variation through the error information carried by error code syndrome measurements. 

\subsection{Classification with Supervised Learning}

Assume we have some circuit $C$, where $C$ is a high-level description of a circuit that has not yet been transpiled to run on any particular quantum backend. Assume also that $C$ includes error-correction operations.

Finally, assume the provider allows users an enrollment phase, wherein the user has trusted access to the provider's backends and can run arbitrary circuits. (Any entity the user trusts can complete this phase; it does not have to be the user directly.) The enrollment phase is a period where we assume we can run circuits on backends with known labels to gather training data.

\paragraph{Training Process} This encompasses the enrollment phase.
\begin{enumerate}
    % \item Compile $n_b \cdot n_m$ versions of $C$, each labeled $C_{b,m}$, where $b \in [1,n_b], m \in [1,n_m]$, $n_b$ is the number of backends, and $n_m$ is the number of mappings. That is, compile the circuit once for every possible combination of backend and random permutation representing a mapping. (TODO: this could probably be worded better.)
    % Compile $n_b \cdot n_m$ versions of $C$, each labeled $C_{b,m}$, where $b \in [1,n_b], m \in [1,n_m]$. That is, compile the circuit once for every possible combination of backend and random permutation representing a mapping.
    \item Transpile $C$ once for each backend $i$ to produce a set of transpiled circuits $\qty{C_i}$.
    \item For each backend, run the circuit transpiled for that backend multiple times (i.e. for many shots) and collect the syndrome measurements (classical bit strings).
    \item Feed the syndrome measurements into a classification model (we used a neural network as described in \cref{sec:supervised-verification}). \label{step:train-model}
    \end{enumerate}
    
\paragraph{Verification Process} The enrollment phase is over, and we no longer assume trusted access to the backends.
    \begin{enumerate}
    \item At a later time, submit circuit $C$ to the provider, requesting that it be run on a specific backend $j$. Collect syndrome measurements. \label{step:verify}
    \item Feed syndromes into the classifier to infer label $k$; verify that $j=k$. Otherwise the user assumes the provider acted dishonestly and ran the user's circuit on a backend other than the one the user requested.
\end{enumerate}

The process of gathering QEC syndrome bit strings for classification is illustrated in \Cref{fig:training}, where we show a circuit transpiled for and run on three backends $A,B,C$. The error syndromes are then collected and run through classification. Note that, in general, we may choose to use multiple rounds of error correction for classification as shown in \cref{fig:training}.

The procedure as described may allow for the provider to transpile the circuit. We will assume that the provider transpiles the circuit similar to how it was transpiled in the enrollment phase. However, we show in \cref{sec:vary-mappings} that our method is robust to variations in how qubits are mapped from logical qubits to hardware qubits (a significant portion of transpilation).

\subsection{Training over Time and over Calibrations}
Various physical changes occur in the environment around qubits that necessitate recalibrating control devices. These changes might include temperature fluctuations, changes in an ambient electromagnetic field, etc. \cite{proctor2020drift,krantz2019superguide} Quantum computing providers periodically take their devices offline, perform calibration measurements, recalibrate their control hardware, and then come back online \cite{ibm2024calibration}. This process may partially diminish our ability to accurately classify backends because it changes the distribution of observed errors. Gate errors specifically are likely to change since the control pulses have been recalibrated \cite{proctor2020drift}. 

To mitigate this, the user may train on error syndromes gathered over multiple calibration cycles. We demonstrate in \cref{sec:stability} that doing so can lead to increased classification accuracy.

Previous quantum fingerprinting schemes overcome this problem in various ways. Mi~et~al.~\cite{mi2021short} use idle tomography-based \cite{blume2019idle} fingerprints, which captures information about how qubits interact with their environment while idle, and doesn't depend on gates. This comes at the expense of needing to run long, dedicated circuits. Smith~et~al.~\cite{smith2023fast} propose using qubit resonant frequencies, which they write are relatively stable. However, this comes at the expense of either requiring provider-given metadata about qubits (their resonant frequencies) or needing to execute a frequency sweep on every qubit, which is also expensive.

% \begin{figure}
%     \centering
%     \includegraphics[width=\linewidth]{figures/training-time.png}
%     \caption{Because of frequent and periodic backend calibrations, the error syndrome distributions change over time. These changes hurt our model performance. To make the models robust to calibrations, we train over at least two calibrations. We speculate that this gives the model the chance to learn error patterns that are stable over time, and ignore patterns that change with calibrations. We then test on data several calibrations out (around a week) to ensure robust performance even when testing within a calibration cycle that has no training data. TODO: This is too small. Fix diagram.}
%     \label{fig:training-time}
% \end{figure}

\section{Experiments and Results}
\label{sec:results}

\subsection{Supervised Verification}
\label{sec:supervised-verification}
To provide a proof of concept of our proposed method, we test a PUF-like protocol with an enrollment phase. We choose a simple circuit that can be error-corrected with realistic error codes on current quantum computers. Error codes greatly amplify the size of circuits because many physical qubits are needed to represent a single logical qubit, so it is only feasible to test small circuits on current hardware.

\begin{figure}
    \centering
    \includegraphics[width=0.7\linewidth]{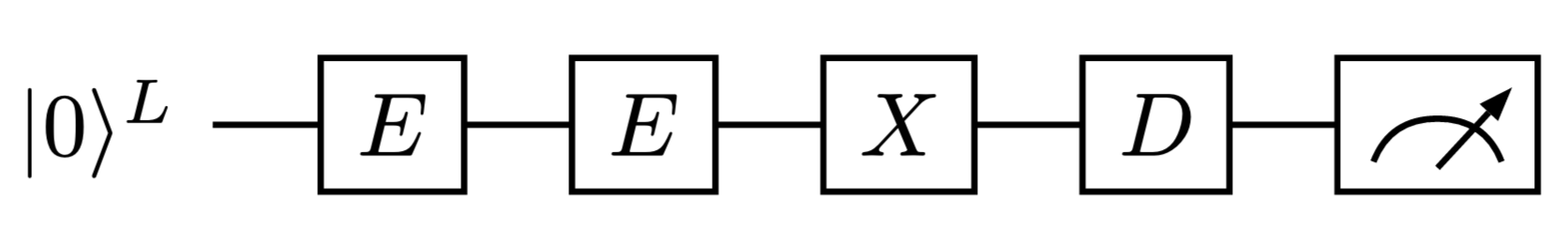}
    \caption{Circuit to test enrolled verification. $E$ and $D$ both represent the "stabilize" operation. Applying stabilize twice is effectively an encoding operation \cite{fowler2012surface}.}
    \label{fig:rep-surface}
\end{figure}

Our chosen ``representative'' circuit --- in that the circuit represents whatever the user would be running --- is shown in \cref{fig:rep-surface}: a simple $X$ gate and then a measurement (with error encoding and decoding operations on either side of the $X$ gate).

The primary goal of this paper is to demonstrate the feasibility of using QEC error syndromes as a basis for fingerprinting for verification. To that end, we demonstrate that various factors \emph{do not} degrade our results. That is, we show that our method generalizes over several key factors described below:

\begin{description}
    \item[Mapping] The function that determines which physical qubits will represent the virtual qubits in the high-level circuit.

    \item[Starting state] Whether a state is, for example, in $\ket{0}$, $\ket{1}$, or $\ket{+}$ may affect measured error syndromes.

    \item[Stability] Classification may perform better within the same calibration cycle, but it is more realistic to consider a model that may be used over multiple days or even weeks without retraining. Experiments may target intra- or inter-calibration performance. 

    \item[Optimization levels] High-level circuits are transpiled before they can run on a backend, and varying the degree of optimization may affect the measured error syndromes.
\end{description}

We additionally test three error correcting codes to show the robustness of the method to different codes:

\begin{description}
    \item[Shor code] \cite{shor1995code} is the first general error-correcting code that handles both bit-flips and phase-flips. It is well-studied and useful pedagogically.

    \item[Steane code] \cite{steane1996code} is an advancement over the Shor code that can make the same general error-correcting guarantees but using fewer qubits (7 instead of 9). We study Steane because it is more practical than Shor but still relatively simple to understand and implement.

    \item[Surface codes] \cite{fowler2012surface} are a family of codes that are expected to be good candidates for real-world use due to their good locality (operations can be done only on neighboring qubits) and high error threshold, meaning that the underlying physical qubits can be relatively noisy and the code will still reduce the error. Surface codes use a relatively large number of qubits, which constrains the representative circuit (\cref{fig:rep-surface}) we use to test the surface code. 
\end{description}

Altogether, the many factors under consideration result in a large matrix of possible experiments to run. Presented in the following sections is a sampling of that matrix. Each section will specify which variables were held constant and which were varied.

% The surface code is parameterized over a distance $d$, which for our purposes simply controls how many errors the code can detect, and also the size of the code. Higher values of $d$ correspond to being able to correct more errors at the expense of a larger code (i.e.~more physical qubits per logical qubit). We tested both $d=3$ and $d=5$ for the surface code, and found that $d=5$ yielded higher classification accuracy;  \dan{watch vague statements like this.  More accurately identifying the back-end?  Just "better" as a QEC in practice?}. This is intuitive since the larger code means more crosstalk and also better access \dan{again vague} to logical error information.

For our experiments, we generally take a single set of syndrome measurements (together the set is one classical bit string) from one circuit execution (shot) to be one feature vector. Quantum circuits are typically run for many shots, or many samples of the measurement distribution. We then run inference on each shot's feature vector individually, and aggregate by taking the mode of all inferred labels. Intuitively, because we have access to the output of many shots for validation, the overall accuracy of inference will be higher than if we had access to only a single shot. We are validating using more data --- more samples from the distribution representing a specific backend's execution of the circuit.

We will then discuss two kinds of accuracy statistics: one for a single shot, and one for the aggregation of many shots. The important result is the accuracy we can achieve with a realistic number of shots. This is discussed in detail in \cref{sec:classiyfing-backends}, but the takeaway is that so long as we can perform better than random guessing, we can scale to high accuracy using a realistic number of shots (>99\% within 500 shots). Unless otherwise indicated, we will therefore report results after \cref{sec:classiyfing-backends} in terms of single-shot accuracy and show that it is higher than random guessing. So although single-shot accuracies are low ($\approx 30\%$), they can scale to high accuracies when aggregated together (>$99\%$).

\subsection{Classifying Backends} %with constant mapping
\label{sec:classiyfing-backends}

\begin{figure}
    \centering
    \includegraphics[width=1\linewidth]{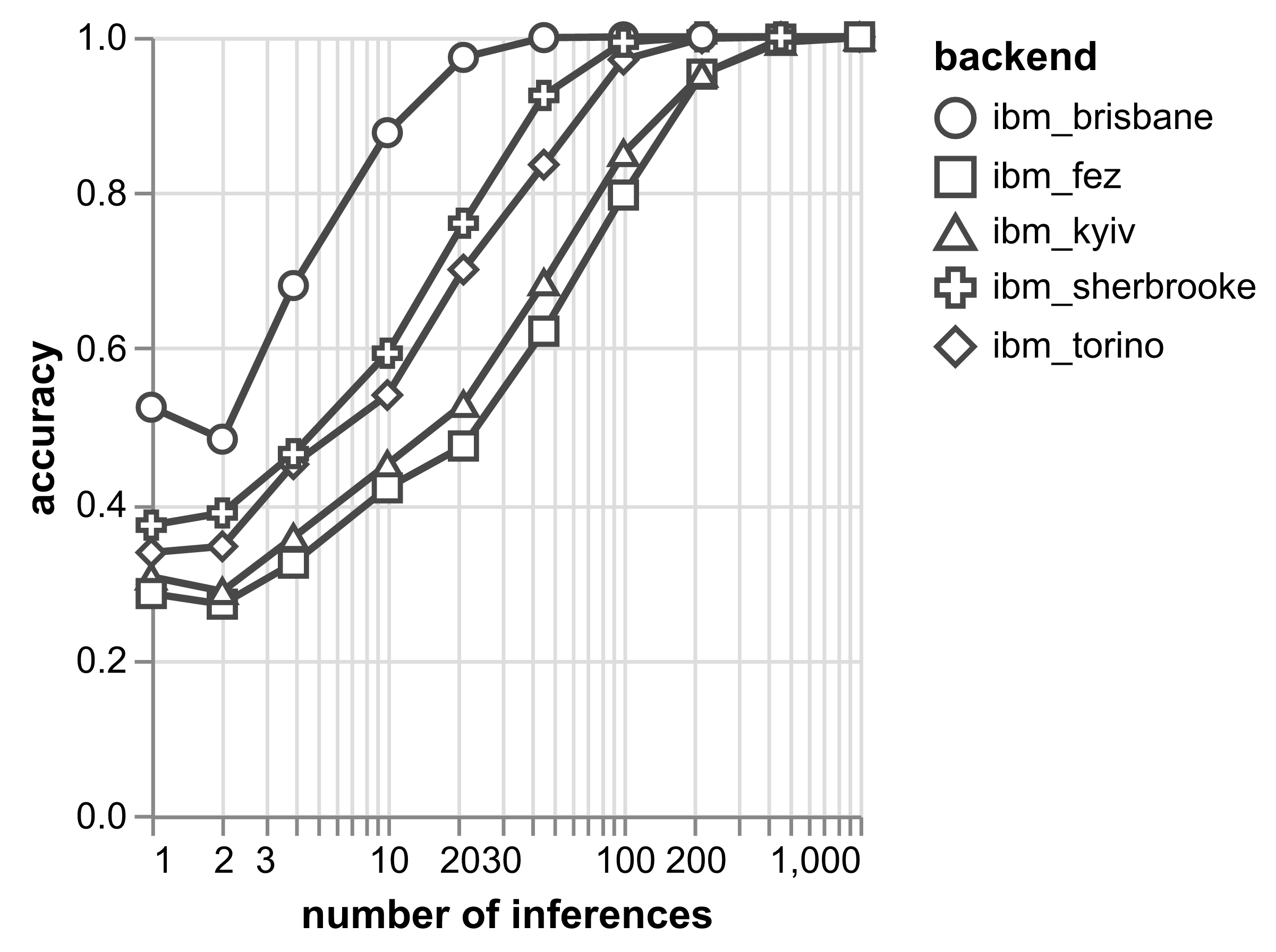}
    \caption{Performance of backend-classification as number of shots increases. As intuition might suggest, more shots (samples) affords better performance. After 500 shots, all backends test at above 99\% accuracy. The trend suggests that we can push this higher with more shots per circuit.}
    \label{fig:perf-vs-n}
\end{figure}

Our fundamental result demonstrates the feasibility of developing a fingerprinting method to identify a quantum backend's identity based on error syndrome measurements. We transpile our simple representative circuit once per five backends from: \texttt{ibm\_brisbane}, \texttt{ibm\_sherbrooke}, \texttt{ibm\_kyiv}, \texttt{ibm\_fez}, and \texttt{ibm\_torino}. We fix a single mapping per compiled circuit, and run these circuits several times. We then train a classifier on the resulting syndrome measurements to predict which backend produced a given syndrome string.

Specifically, we start with the simple circuit shown in \cref{fig:rep-surface} with a logical starting state of $\ket{0}$, using the surface code for error correction. We then transpile this circuit for the five backends listed above, setting \texttt{qiskit}'s mapping strategy to \texttt{trivial}. From \texttt{qiskit}'s documentation \cite{trivial-docs}, the \texttt{trivial} layout is a ``pass for choosing a Layout of a circuit onto a Coupling graph, using a simple round-robin order,'' which means qubits are mapped from physical to virtual in simple ``increasing order''. We run each of these circuits over several days, and then train on a random sample of about 540,000 shots (about 108,000 per backend), where the classification labels are the backend that produce the shot (the output of one shot corresponds to one feature vector). We use a neural network (MLP) \cite{sklearn,pytorch} with one hidden layer of size 128. 

The results are shown in \cref{fig:perf-vs-n}. Performance on classifying a single syndrome bit string (from a single shot of a circuit) is at least about 30\%. However, this is still above random guessing (which we would expect to be about 20\% with five backends), so we see that there is signal to be extracted. This signal can be boosted by taking more syndrome measurements. User circuits are typically run thousands of times, so we will naturally have thousands of samples from the syndrome measurement distribution on which to run inference.

To aggregate the results of multiple inferences, we simply take the mode of those results. We tried other aggregation methods, including batching many shots' outputs into one vector by taking the element-wise mean of each qubit. Taking the mode gave us the best accuracy over multiple shots.

% \dan{Need more details on experimental process, i.e., how many shots were the trained classifiers run on?  What is the method of training?  How are multiple shots used to select backend - closest mode (L1 or L2 norm?) etc.  What mappings were used?  Start states?  Which ECC?}

We see from \cref{fig:perf-vs-n} that our inference accuracy scales with the number of shots. Our current method seems to scale logarithmically, though in practice this is not a problem because we reach high accuracy (>$99\%$) within around 500 shots, which is likely fewer than most business circuits will require. For example, variational quantum eigensolver (VQE) algorithms are expected to be useful for a wide range of tasks including quantum chemistry, material science, and finance \cite{mcardle2020qchemistry,herman2023finance}. Quantum chemistry applications typically require a mean error of less than $1.6 \times 10^{-3}$ --- so-called ``chemical accuracy'' \cite{mcardle2020qchemistry}. Recent efforts \cite{arrasmith2020shots} to decrease the number of necessary shots still require, for example, roughly $10^5$ shots to achieve chemical accuracy in estimating the ground state of $\mathrm{H}_2$, a small molecule. K{\"u}bler~et~al.~\cite{kubler2020vqe} test adaptive optimizers to reduce the required shot count in variational quantum algorithms and do not report results fewer than 1000 shots for any error rate. We believe it is a safe assumption that users will predominantly run algorithms that execute more than 500 shots in total.

These experiments demonstrate that information about backend identity is contained in error-correction syndromes. By extracting this information from error syndrome measurements, we develop the basis for a system to authenticate quantum backends to users. 

\paragraph{Calibrating Models}
When using a uniform cost function per backend, we find that the models we train skew heavily toward one class over another. For example, when training the model that generated \cref{fig:perf-vs-n}, we find that a uniform cost leads to \texttt{ibm\_kyiv} and \texttt{ibm\_fez} taking nearly 10,000 shots to converge to over 90\% accuracy; other classes take only around 100 shots. To demonstrate that we can achieve high classification across all backends simultaneously, we manually modify the class weights for the model to boost the importance of \texttt{ibm\_kyiv} and \texttt{ibm\_fez}. This leads to other backends taking slightly longer to converge --- 500 shots instead of 100.

\subsection{Robustness to Error-Correcting Codes}
\label{sec:vary-codes}

\begin{table}[]
\begin{tabular}{@{}llll@{}}
\toprule
Code & Accuracy & FPR & FNR \\ \midrule
Shor & 0.89      & 0.04  & 0.01   \\
Steane & 0.72 & 0.12 & 0.28         \\
Surface ($d=5$) & 0.83  & 0.08  & 0.17   \\ \bottomrule
\end{tabular}
\caption{Summary of backend classification performance. The model is trained on days 1 and 2, and tested on day 10. Inference is run over aggregated 40 shots.}
\label{tab:summary-backend}
\end{table}

We further validate our findings by demonstrating their robustness to various parameters of interest, starting here with the selection of error correcting code. We test three different codes: Shor, Steane, and surface codes. Our high level circuit for the surface code is again the same as in \cref{fig:q-rep-code}. For Shor and Steane we are able to run larger logical circuits because they require fewer qubits, so we run a CNOT on logical $\ket{1,0}$ instead of an $X$ on logical $\ket{0}$ (otherwise the circuits are the same). We again train a classifier to infer on which backend the circuits were run. Each circuit uses a fixed starting state ($\ket{1,0}$ or $\ket{0}$) and layout optimization level (\texttt{trivial}). Our results are summarized in \cref{tab:summary-backend}, which shows that for all three codes tested, we achieve classification performance significantly higher than random guessing (greater than $70\%$ compared to 33\% expected for random guessing). 

The accuracies reported here are for feature vectors derived from 40 shots aggregated by taking their mean. We see similar accuracy at around 40 shots from \cref{fig:perf-vs-n}, which is as expected because that figure represents the surface code.

\subsection{Robustness to Varied Mappings}
\label{sec:vary-mappings}

%TODO?
% To test mapping classification, we began by choosing a random permutation to act as an initial ``qubit mapping'' --- essentially a starting state for the compiler to allocate certain physical qubits to certain logical qubits in the circuit. An arbitrary permutation is unlikely to be a valid mapping for a circuit, so we let the compiler modify our random initial mapping to work with the constraints of the chosen backend. We limit these modifications by using low optimization levels (tested using the \texttt{trivial} parameter in qiskit). Otherwise, two distinct mappings may be optimized to the same third, lower-cost mapping.

Although IBM and other providers currently allow users to specify low-level details of circuit execution, such as the mapping from logical to physical hardware qubits, we want to verify that our method will continue to be useful if future interfaces are higher level and abstract away these details. We show that our method allows consistent classification of syndrome measurements produced by circuits transpiled to different mappings, even mappings that were not in the training set.

To demonstrate this, we first take our high-level circuit (\cref{fig:q-rep-code}) and transpile it exactly once for each backend. We allow \texttt{qiskit} to choose an arbitrary mapping. For each transpiled circuit, we then generate sixteen random re-mappings of that circuit onto the same hardware. Every quantum backend has some connectivity graph that determines which qubits can directly interact with each other. Any re-mapping we generate must obey this connectivity graph. To find suitable re-mappings, we use the following process:

\begin{enumerate}
    \item Extract the used connectivity subgraph from the transpiled circuit output by \texttt{qiskit}. That is, since we are only using a subset of the available hardware qubits for our circuit, we find the \emph{subgraph} of the full backend connectivity graph induced by the set of qubits we actually use for the computation.

    \item Find subgraphs in the full backend connectivity graph that are isomorphic to the subgraph induced by our used qubits. For this experiment, we find sixteen isomorphic subgraphs per backend.

    \item Manually edit the \texttt{qiskit} object to use the hardware qubit indices from our new isomorphic subgraph. (Repeat for each of the sixteen found isomorphic subgraphs.)
\end{enumerate}

This approach allows us to gather syndrome data for multiple qubit layouts on a single backend without changing the underlying connectivity. We go through this process to be sure that the only difference between the sixteen transpiled circuits per backend is their mappings, instead of possible confounding variables introduced by the transpilation process.

\begin{figure}
    \centering
    \includegraphics[width=1\linewidth]{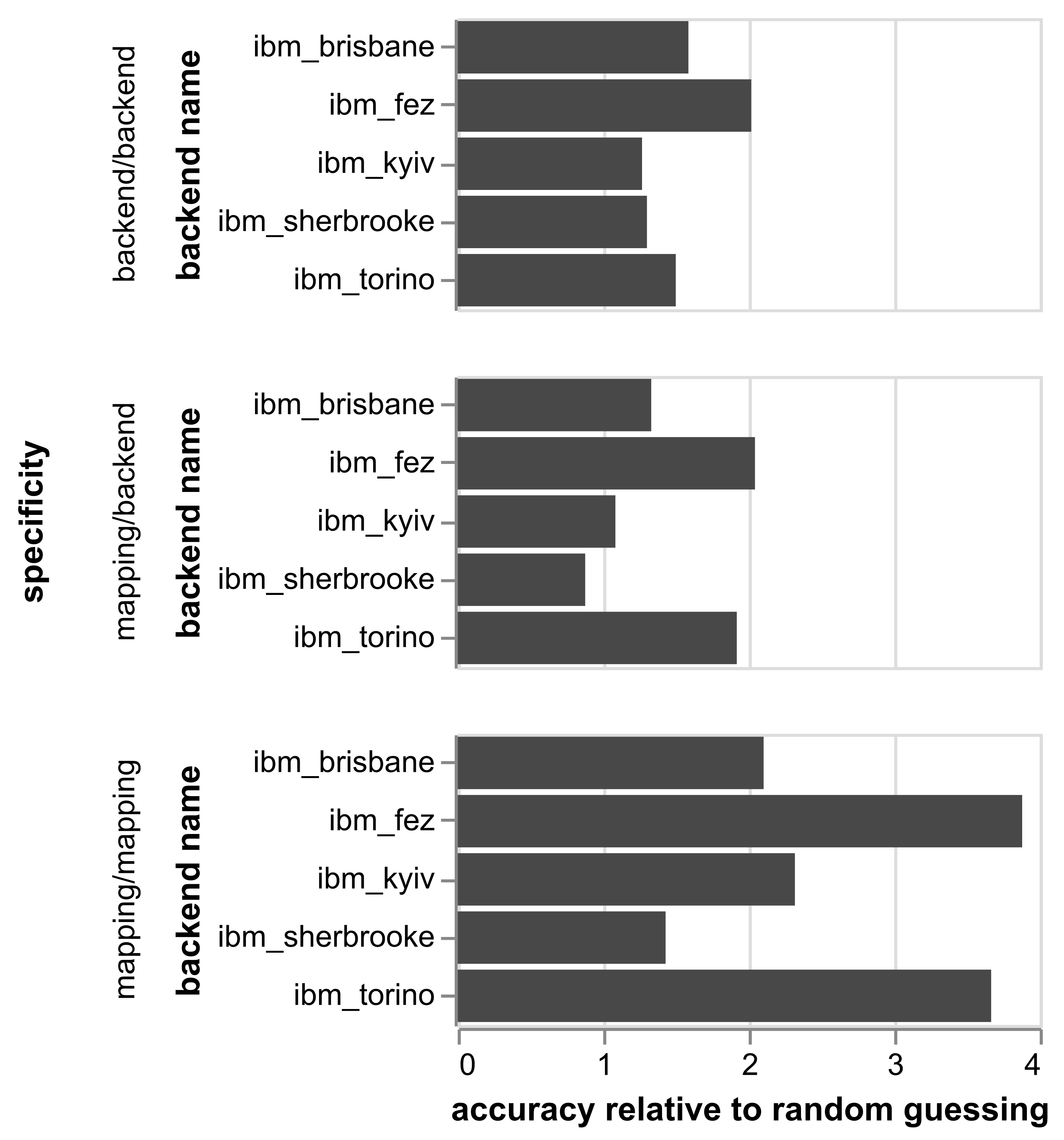}
    \caption{Our model performs well even when required to train over multiple mappings. Note that random guessing for the top two plots is 20\%, but only 1.3\% for the bottom plot. All three instances perform significantly better than random guessing. Mean accuracies were 31\%, 29\%, and 3.6\% from top to bottom. The plots are split by ``specificity'', which is read as ``training/inference'', e.g.~the plot labeled ``mapping/backend'' refers to the experiment where the model is \emph{trained} with one label per mapping (per machine), and \emph{inference} is run only requiring the model to accurately guess the backend to be considered correct.}
    \label{fig:mapping-perf}
\end{figure}

We ran this process on the circuit described in \cref{fig:q-rep-code} with a distance-3 surface code on a starting state of $\ket{0}$. Our results are shown in \cref{fig:mapping-perf}, which is divided into subplots by a category we call {\em specificity}, which refers to whether the labels for training or classification are the \emph{backend} identity or the \emph{backend-mapping} pair.

We write ``backend/backend'' to describe a model trained and tested on only backend identities. That is, the only information in the labels is the backend, and no information about what mapping is used for that shot. The first plot in \cref{fig:mapping-perf} shows this case. There are several mappings per backend in this data, and we achieve a mean accuracy of 31\%. This is significantly higher than random guessing (20\%), so we find positive signal (we can scale the accuracy higher with more shots as shown in \cref{fig:perf-vs-n}). This result demonstrates that we can still accurately predict backend identities even when there are multiple mappings per backend under consideration.

We write ``mapping/mapping'' to describe a model trained and tested on backend-mapping pairs. Each backend-mapping pair is identified as a unique class to the model. This case is shown in the bottom plot in \cref{fig:mapping-perf}. The performance seems low at only 3.6\% mean accuracy, but it is actually nearly $2.8 \times$ more accurate than random guessing, which we would expect to be 1.3\% (there are 80 pairs of 16 mappings over five backends each). Though it is not the primary focus of this paper, these results indicate error syndromes carry information not just about the backend, but about which specific hardware qubits were used in a computation. This is potentially useful for building a ``strong PUF'' \cite{gassend2002silicon} since the space of potential mappings is large if the quantum computer is significantly larger than the circuits it is running (i.e.~there is room to move the circuit around and find new isomorphic subgraphs).

Finally, we write ``mapping/backend'' to describe a model trained on backend-mapping pairs, but tested only on backend inference. That is, the training data have labels that look like \texttt{(ibm\_brisbane, mapping\_5)}, but we judge accuracy by considering only the backend. We see from the middle plot in \cref{fig:mapping-perf} that our performance is comparable to the ``backend/backend'' case; the mean accuracy is 29\% (again significantly better than random guessing at 20\%). This case is interesting because even though the model does not know about backends as a separate concept from mappings, we see that its confusion between mappings is usually contained within the same backend. So syndromes from differently mapped circuits within the same backend are more similar than those from different backends.

\subsection{Robustness to Starting States and Layout}
\label{sec:vary-states-layout}
\begin{figure}[h]
    \includegraphics[width=\linewidth]{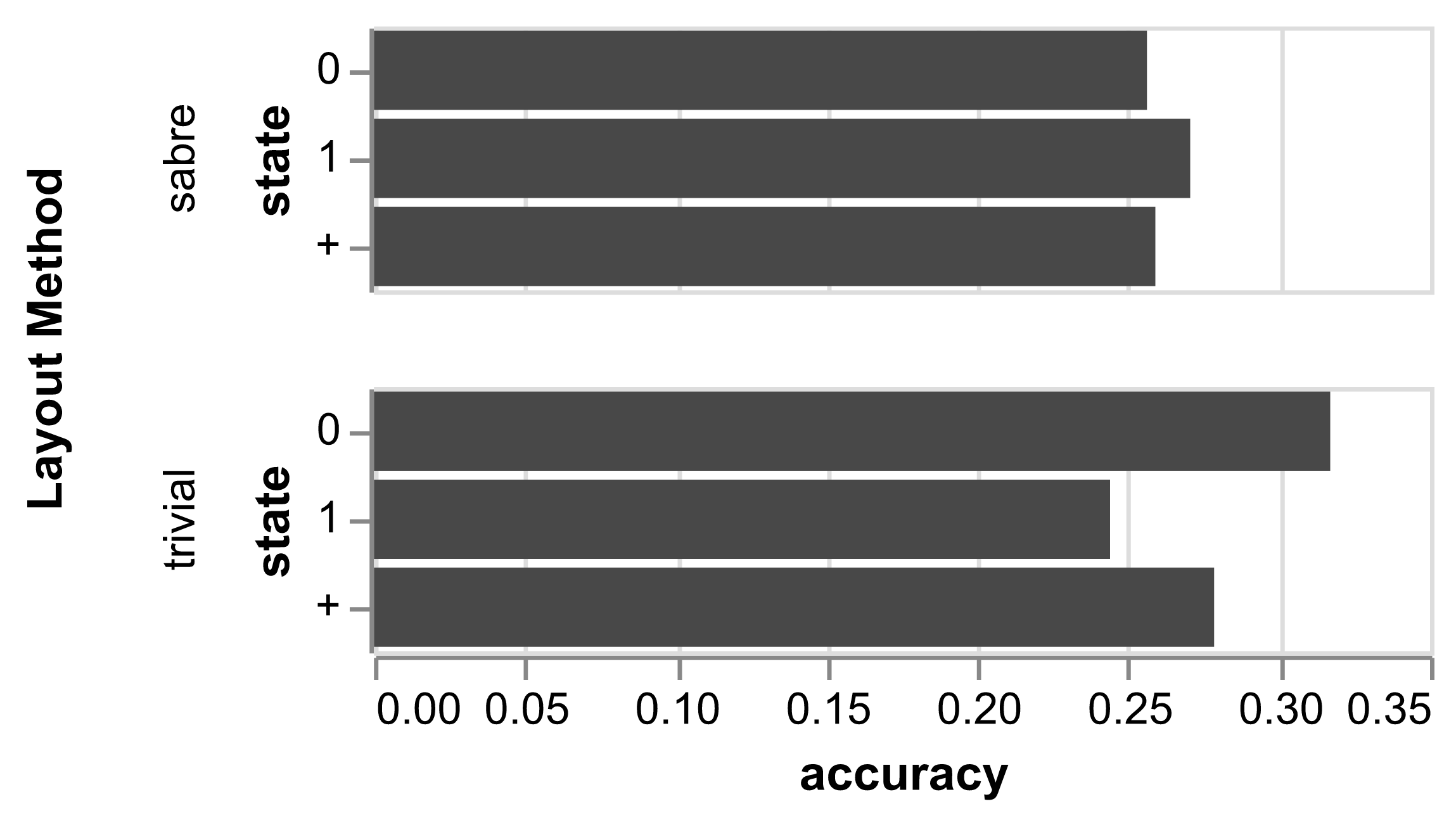}
    \caption{Classification performance does not significantly vary with optimization levels. In this result, we refer to layout optimization, where the mapping between logical qubits in a circuit and physical qubits on the device is arranged to minimize expected error.}
    \label{fig:layout}
\end{figure}

Different quantum states are susceptible to different errors; the simplest example is that the (conventionally) higher-energy state $\ket{1}$ will experience relaxation errors and decay into $\ket{0}$, while $\ket{0}$ does not as readily flip to $\ket{1}$. Further, a flip from $\ket{0}$ to $\ket{1}$ is not a decay, but rather absorption of energy from the environment \cite{krantz2019superconducting}, so different states can be affected by different error channels.

To test our method's resilience to varied starting states, we create circuits the same as in \cref{fig:q-rep-code} except with different logical starting states. In addition to $\ket{0}$, we test both $\ket{1}$ and $\ket{+}$. As before, we execute these three circuits across all five backends and feed the syndromes into a classifier. We demonstrate in \cref{fig:layout} that all three starting states result in similar accuracy, averaging around 28\% after the circuits undergo \texttt{trivial} mapping optimization, and 26\% under the more realistic (in the sense that this is what users will actually do) SABRE optimization algorithm \cite{li2019sabre,zou2024lightsabre}.

The results in \cref{fig:layout} are split by the layout optimization method to demonstrate that classifying by error syndromes is also robust to the compiler's layout algorithm. The layout algorithm aims to reduce the number of SWAP gates and thereby lower the overall error of the circuit \cite{li2019sabre}, so we might wonder if higher optimization levels remove error and lower the performance of our classification method. However, the mean accuracies (just described) are nearly identical, and both are significantly greater than random guessing, so we conclude that our fingerprinting method is robust to changes in both starting state and optimization method (for layouts).

\subsection{Stability over Time}
\label{sec:stability}

Qubits are sensitive to changes in their environment \cite{proctor2020drift}, and the control systems operating on qubits have to be tuned precisely. Therefore, operators of quantum computers have to periodically recalibrate their systems to account for the drift in physical qubit properties over time \cite{ibm2024calibration}. This recalibration will at least change the prevalence of gate errors and therefore the distribution of measured syndromes.

\begin{figure}
    \includegraphics[width=\linewidth]{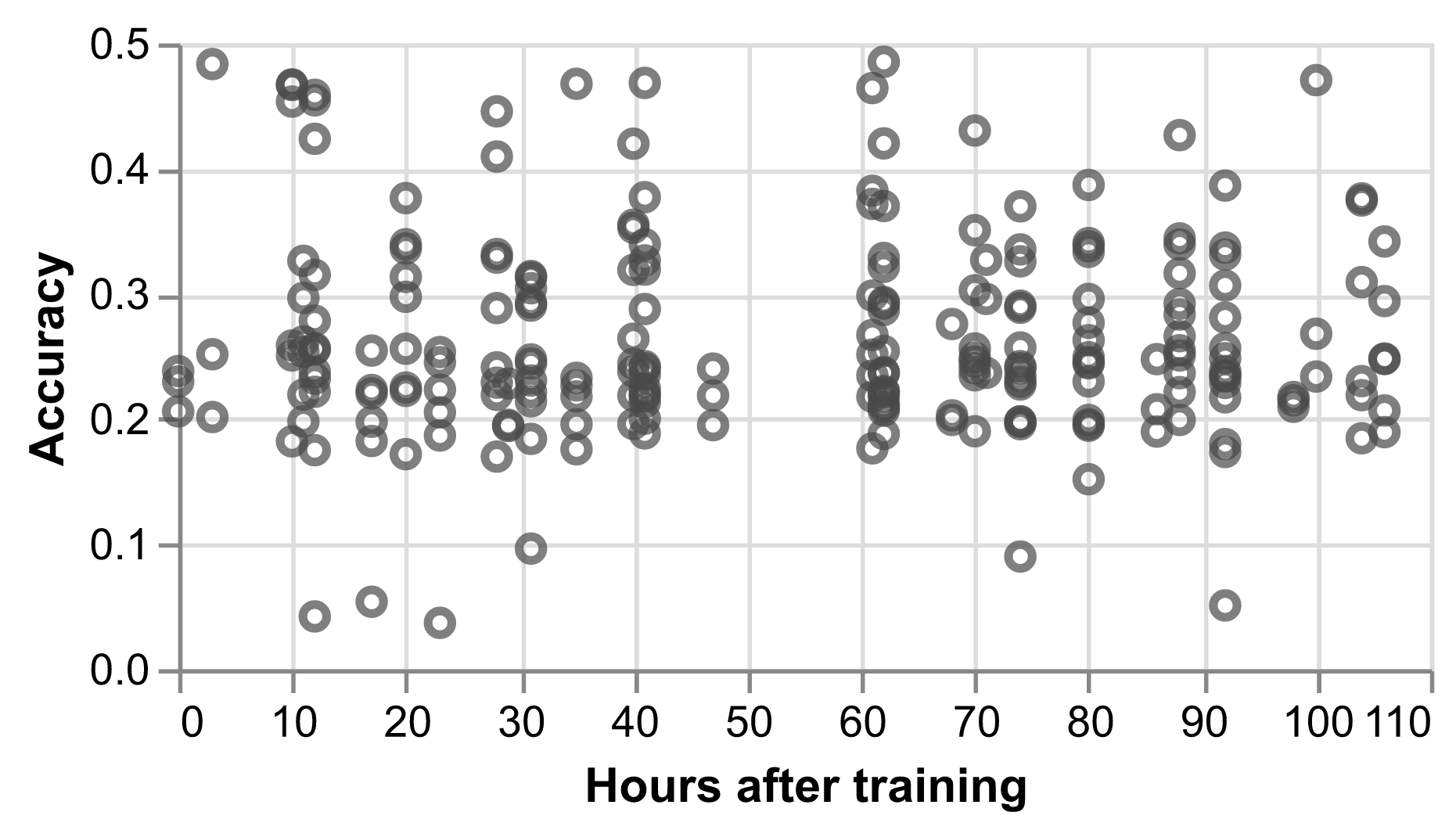}
    \caption{Classification performance does not significantly degrade over time. We prepare one circuit per starting state per backend, and run these circuits repeatedly over 100 hours. We find that mean performance is approximately constant over this time period. All three tested starting states show similar mean accuracy (shown explicitly in \cref{fig:layout}. }
    \label{fig:decay}
\end{figure}

We demonstrate that our models are robust to these changes by training on data collected within one calibration cycle, and then testing the models on the same circuits run after several calibration cycles over the next four days. We additionally ran these tests for three different starting states, and all performed similarly, so the results are all plotted together in \cref{fig:decay}. Over four days we see no apparent decline in model quality, even going from testing within the same calibration cycle to testing between different calibration cycles in the first hours of the experiment.

Additionally, the results from \cref{tab:summary-backend} indicate that our method is stable over time periods of at least about ten days. The experiment described by \cref{tab:summary-backend} consisted of training on day 1 and testing on day 10. We measured accuracies of at least 70\% for all codes with as few as 40 shots (aggregated into feature vectors with element-wise mean).

Finally, we test a simple method of increasing model accuracy when testing across multiple calibration cycles with the same model. We hypothesize that providing multiple cycles of training data would allow the model to learn patterns that are maintained across calibration cycles. To test this, we compile a simple CNOT circuit using the Steane code and execute it over three days: day 1, day 2, and day 7. When trained on syndromes from day 1 and tested on syndromes from day 7, we observe an accuracy of 50\%. When trained on syndromes from both day 1 and 2 and tested on day 7, we observe an accuracy of 64\%. This is a significant increase in accuracy, but it is worth noting that we do not observe the same increase in accuracy in the experiment shown in \cref{sec:stability}. It is possible that we are simply observing the effect of adding more training data.

\subsection{Unsupervised Modeling of Backend Changes }
\label{sec:unsupervised-results}
Requiring an enrollment phase may be a burden on some users or providers. We therefore test an online learning and classification method that measures changes in the distribution of syndrome measurements over time. This method cannot precisely \emph{identify} the backend used to run a circuit, but it can detect \emph{changes} in backend. So long as the provider is honest at least once, the user should detect dishonest behavior.

Since users will not have access to backend labels, we use unsupervised learning methods to train our model. We will then verify these against backend labels only for the purpose of evaluating our unsupervised model.

We first show that there is a demonstrable difference between the measurement distributions between two circuits from the same backend versus two circuits from different backends. This property is what will allow us to identify a change in backend, which we assume is always a dishonest action by the provider. The following experiment describes a procedure that may be used to detect the difference in these distributions.

For the following steps, assume a fixed starting state of $\ket{0}$ and syndromes measured from the same simple circuit shown in \cref{fig:rep-surface}. We apply only \texttt{trivial} optimization to these circuits. The backends tested are \texttt{ibm\_brisbane}, \texttt{ibm\_sherbrooke}, \texttt{ibm\_kyiv}, \texttt{ibm\_fez}, and \texttt{ibm\_torino}.

\begin{enumerate}
    \item We run the representative circuit on each tested backend several times over two days (ensuring that at least one calibration cycle occurred).

    \item For each circuit execution, we take the average value of all shots, so that the $i$th value in the array is the probability over that circuit's shots that the $i$th bit would measure $\ket{1}$.

    \item For all pairs of circuit executions $A,B$, we form the tuple $(\abs{t_B - t_A}, D(A,B))$, where $D$ is the Euclidean distance function measured over the averaged shots of executions $A$ and $B$. 
\end{enumerate}

\begin{figure}
    \centering
    \includegraphics[width=\linewidth]{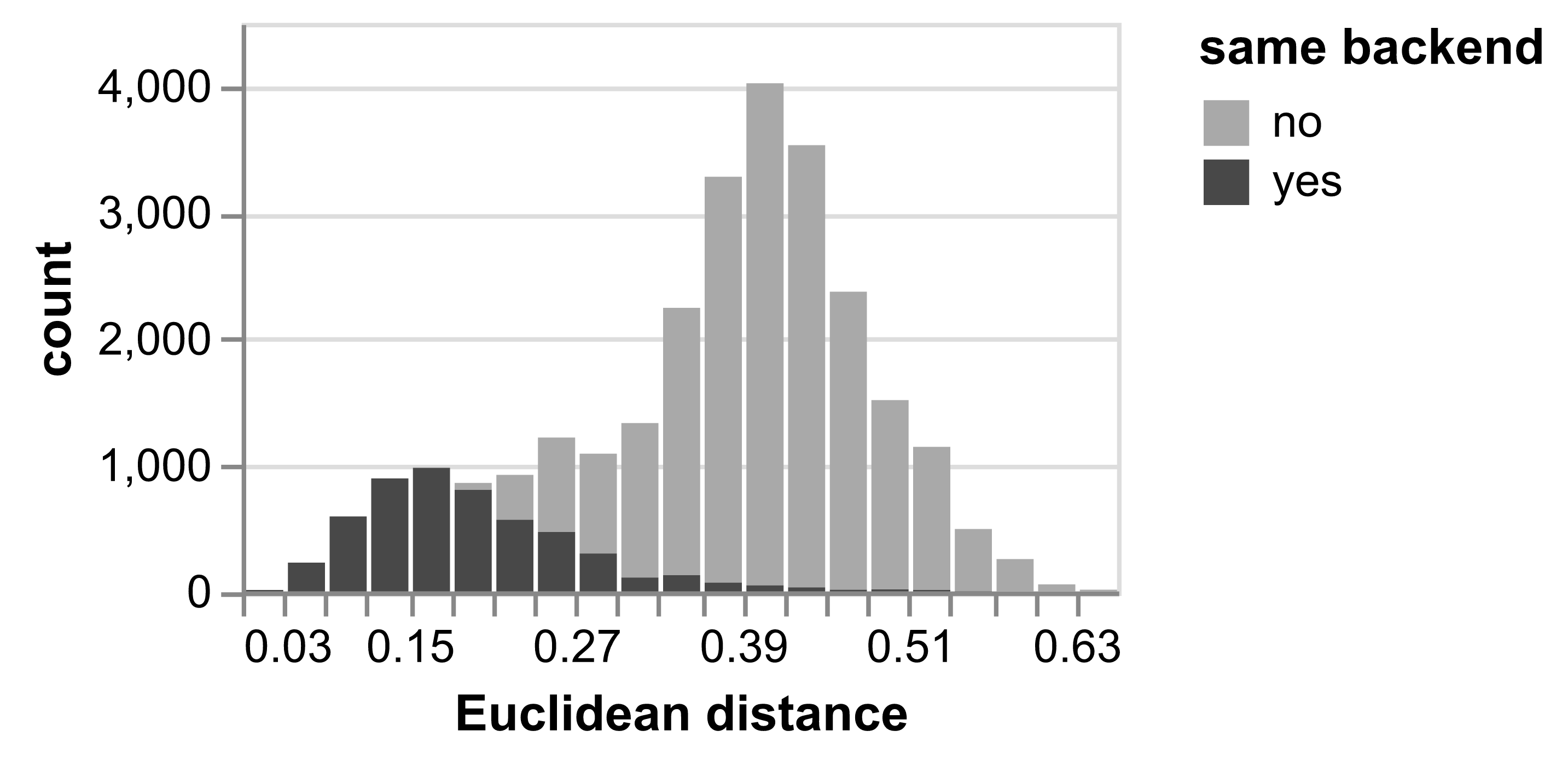}
    \caption{Simple metrics may suffice for untrained online classification. Executions on the same backend are significantly closer under a Euclidean distance metric than executions from different backends. (Not shown, but this effect does not degrade with time, as expected based on our result in \cref{fig:decay}.)}
    \label{fig:unsupervised-euclidean}
\end{figure}

In \cref{fig:unsupervised-euclidean}, we plot histograms of these pairs of distances, separated by whether the pair were from the same machine or different machines. We find that there is no apparent time dependency on the distance metric (as expected based on our results from \cref{fig:decay}), so we exclude time from \cref{fig:unsupervised-euclidean} for clarity. We see a clear separation between same-backend and different-backend job pairs at approximately distance 0.15.

To develop a robust method of verification using unsupervised classification, we use DBSCAN clustering \cite{dbscan}, which takes each job and computes the distance metric with all the other jobs. We vary two hyperparameters to control DBSCAN: $\epsilon$ defines the maximum distance between points to be considered neighbors, and \texttt{min\_samples} sets the minimum number of points required to form a dense region. The idea behind this use of DBSCAN is if the number of pairs with one common job within an arbitrary $\epsilon$ exceeds \texttt{min\_samples}, then we label this a \emph{core point}, and one cluster is formed, representing a backend.

We test our unsupervised model on ranges of parameters to determine what works best for this domain; we use ARI (Adjusted Rand Index) to measure the similarity between the clustering results and true labels, ranging from $-1$ to $1$ (higher is better, 1 is best). We find that for the Euclidean distance metric, by choosing $\epsilon=0.089$ and \texttt{min\_samples} = 5 for the clustering model, we get an ARI of $0.825$.

To see how this model is used, consider the case of a dishonest reseller. The overall goal is to implement a solution for the following scenario: on day $i$, the third-party provider changes the backend without telling the user. In this case, the first-party provider gives the user the error syndromes for each shot. After processing this data by averaging each job’s shot into a single bit vector (as before), we use the optimized DBSCAN cluster and test if the Euclidean distance between this new job and a core point is under the threshold. If that’s the case, then the job belongs to that cluster and we conclude there’s no backend change. If not, this job doesn’t belong to any cluster and there was likely a backend change. This crucially allows us to account for changes to backends that we have not trained on.

% \begin{figure*}[h]
%     \centering
%     \begin{subfigure}[t]{0.5\textwidth}
%         \centering
%         \includegraphics[width=\linewidth]{figures/steane_day1_mapping.png}
%         \caption{No cross-calibration training}
%         \label{fig:perf-mapping-steane-day1}
%     \end{subfigure}%
%     ~ 
%     \begin{subfigure}[t]{0.5\textwidth}
%         \centering
%         \includegraphics[width=\linewidth]{figures/steane_day12_mapping.png}
%         \caption{With cross-calibration training}
%         \label{fig:perf-mapping-steane-day1-day2}
%     \end{subfigure}%
%     \caption{Classification improves when trained across multiple calibration cycles. Shown is the classifier performance using the Steane code to predict mappings both with and without cross-calibration training. Performance is clearly better for \texttt{bris0} and \texttt{bris2}, and overall accuracy improves from 50\% to 64\%.}
%     \label{fig:cross-calibration-stats}
% \end{figure*}

\subsection{Causal Analysis}
\label{sec:causal}

We attribute the classifiability of the devices tested in this experiment to hardware imperfections that result in variations between the noise channels of each device. Metallic surfaces and insulating materials employed in superconducting qubit circuitry often form parasitic amorphous layers during device fabrication and atmospheric exposure \cite{siddiqi2021engineering,murray2021material,krantz2019superguide}. These amorphous defects are unclonable and thus generate average performance variations from device to device that manifest in different sources of noise. We speculate our ability to distinguish between hardware based on outputs of identical circuits may partially be attributed to these effects.

We further investigate which noise sources contribute to these variations by comparing the performance of classifiers trained on data from real hardware to classifiers trained on data obtained from different noise model simulations. By limiting the sources of noise we choose to include in our models and examining the corresponding classification performance, we can deduce which sources contribute to positive classification.

We use two models to emulate different sources of noise: energy relaxation and dephasing (ERaD), and backend emulation. ERaD is the simpler model, taking into account fewer sources of error. The noise is modeled phenomenologically (i.e.~from measurements instead of built from theory and first principles) and characterized by parameters obtained from device calibrations. The parameter values were accessed using the IBM \texttt{qiskit} API, which reports the most recent calibration data.

\begin{description}
        \item[Energy Relaxation and Dephasing (ERaD)] This model considers the time scales over which the qubit population relaxes to its steady state (the ground state for superconducting qubits) and the coherence of a superposition state decays \cite{krantz2019superguide}. The associated noise is characterized by the average times $T_1$ and $T_2$ respectively, and the quantum gate durations $T_g(G_i)$. The argument $G_i$ is a quantum gate from the set of supported basis gates on a backend where $i$ spans the number of gates used in an executed quantum circuit. The model is built using \texttt{qiskit\_aer\allowbreak{}.noise\allowbreak{}.thermal\_relaxation\_error}. A detailed description of this noise model can be found in \cite{georgopoulos2021modeling}.
        %where $G\in\{ECR, RZ, SX, X\}$ in the case of \texttt{ibm\_brisbane}, \texttt{ibm\_sherbrooke}, and \texttt{ibm\_kyiv} and $G\in\{CZ, RX, RZ, RZZ, SX, X\}$ in the case of \texttt{ibm\_fez} and \texttt{ibm\_torino}.
        \item[Backend Emulation] This type of model considers energy relaxation and dephasing noise (like the ERaD model), but also accounts for depolarization and readout errors. Depolarization error is the replacement of the qubit state with the maximally mixed state \cite{mikenike}, while readout error is the erroneous measurement of the qubit state (a significant source of error comparable to 2-qubit gates like SWAP), sometimes caused by crosstalk within the quantum computer \cite{chen2019readout}. These noise channels are characterized by gate infidelities, readout error probabilities, device temperature, and qubit frequency. The model is built from \texttt{qiskit\_aer.noise.\allowbreak{}NoiseModel.from\_backend}. A detailed description of this noise model can be found in \cite{blank2020quantum}.
\end{description}

%, noise exhibits non-Markovian dynamics \cite{smart2022relaxation} which are not captured by our noise models.

\begin{figure}
    \centering
    \includegraphics[width=\linewidth]{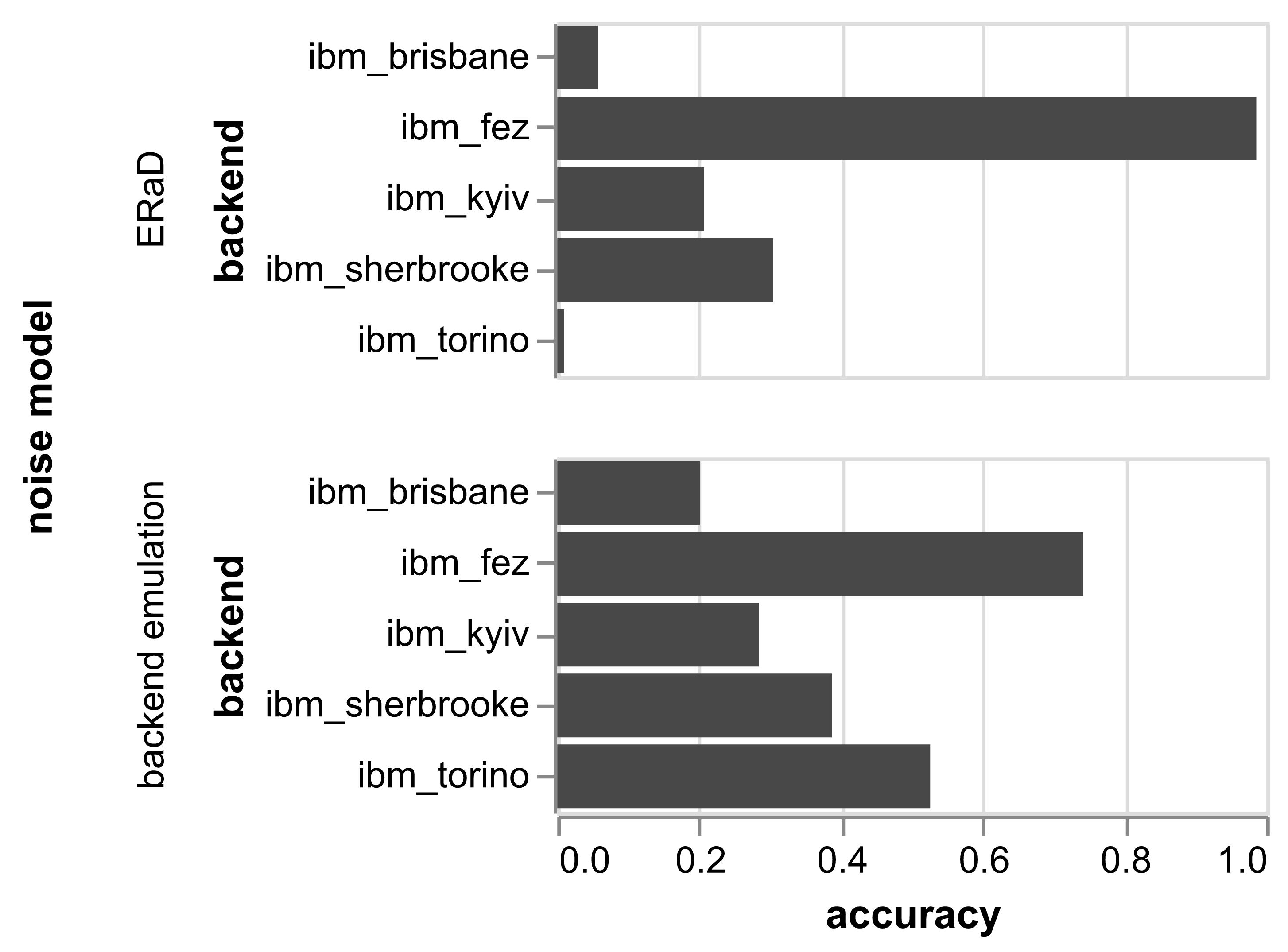}
    \caption{Shown above is the classification accuracy of our model trained on simulated data under two different noise models: ERaD (less complex) and backend emulation (more complex). We achieve higher mean accuracy with backend emulation (42\%) than with ERaD (31\%).}
    \label{fig:simulation}
\end{figure}

A surface code error correction circuit starting in logical state $\ket{0}$ is replicated for emulation using both noise models of all tested backends. Emulation is conducted using the \texttt{qiskit} matrix product state simulator. We conduct 15 emulations per backend which each perform 512 shots of the surface code circuit. We then use the same training process described in \cref{sec:supervised-verification} to create a model from these measured bit strings. The classification accuracy for each error model type and each backend is shown in \cref{fig:simulation}. 

We find that data generated from ERaD is insufficient for emulating classifiability across all tested backends. Data generated from backend emulation yields 42\% classification accuracy on average, while ERaD yields 31\% and failed to classify certain backends (despite our attempts to adjust class weights). This suggests the exclusion of depolarizing channels and state readout error in ERaD obscures the classifiability between backends. Thus, depolarization, readout error, or a combination of other backend emulation-specific noise channels contribute to our extraction of unique device characteristics from error syndromes. 

\begin{figure}
    \centering
    \includegraphics[width=\linewidth]{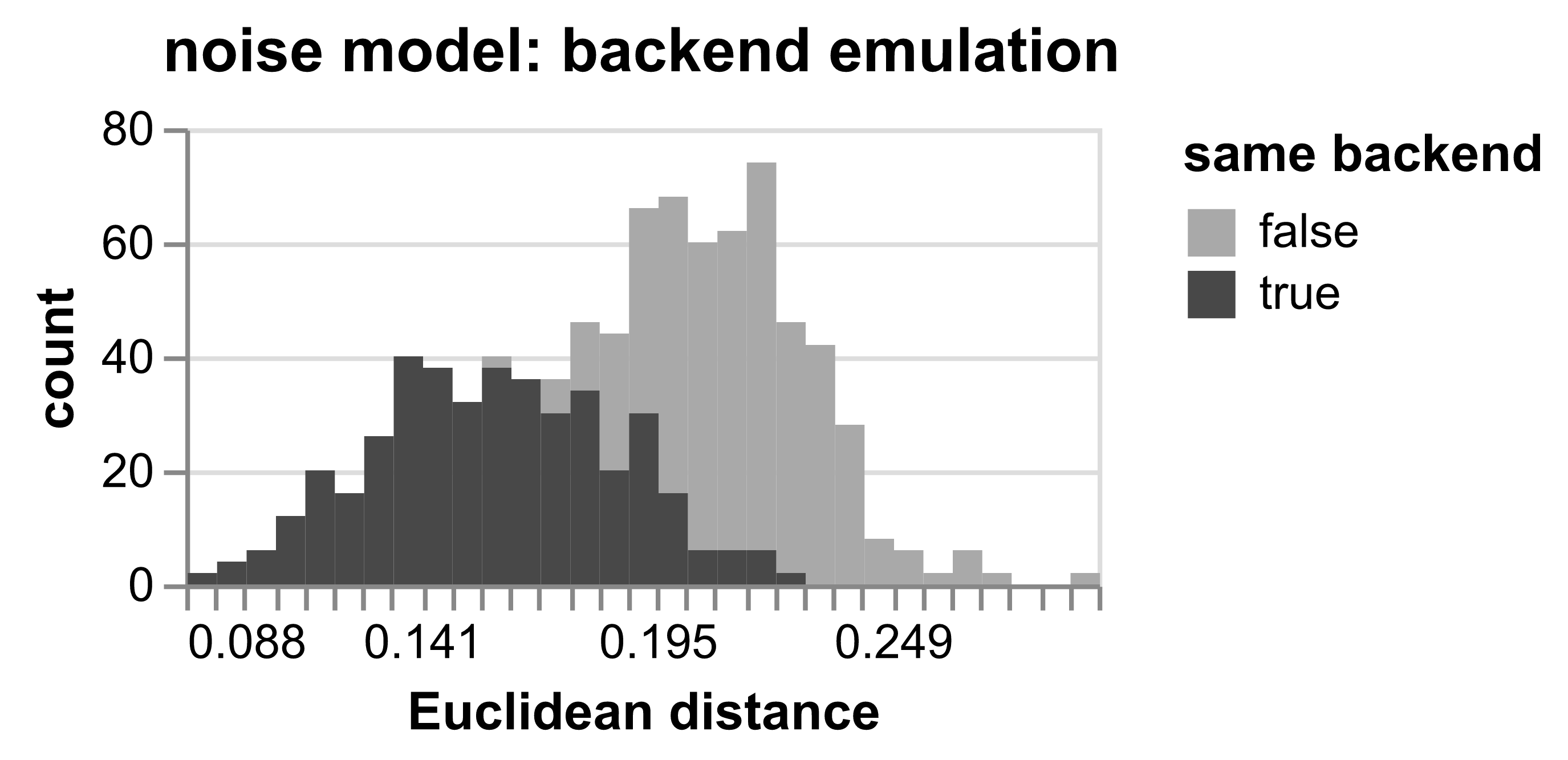}
    \caption{Shown above is a comparison between shots simulated on \texttt{ibm\_brisbane} and \texttt{ibm\_sherbrooke} under the backend-emulation noise model. The plot shows distributions of Euclidean distances between pairs of jobs (all shots averaged down to one feature vector), with the pairs divided into being from either the same or from different backends. We see less separation between same versus different backends than in \cref{fig:euclidean-thermal}.}
    \label{fig:euclidean-complex}
\end{figure}

\begin{figure}
    \centering
    \includegraphics[width=\linewidth]{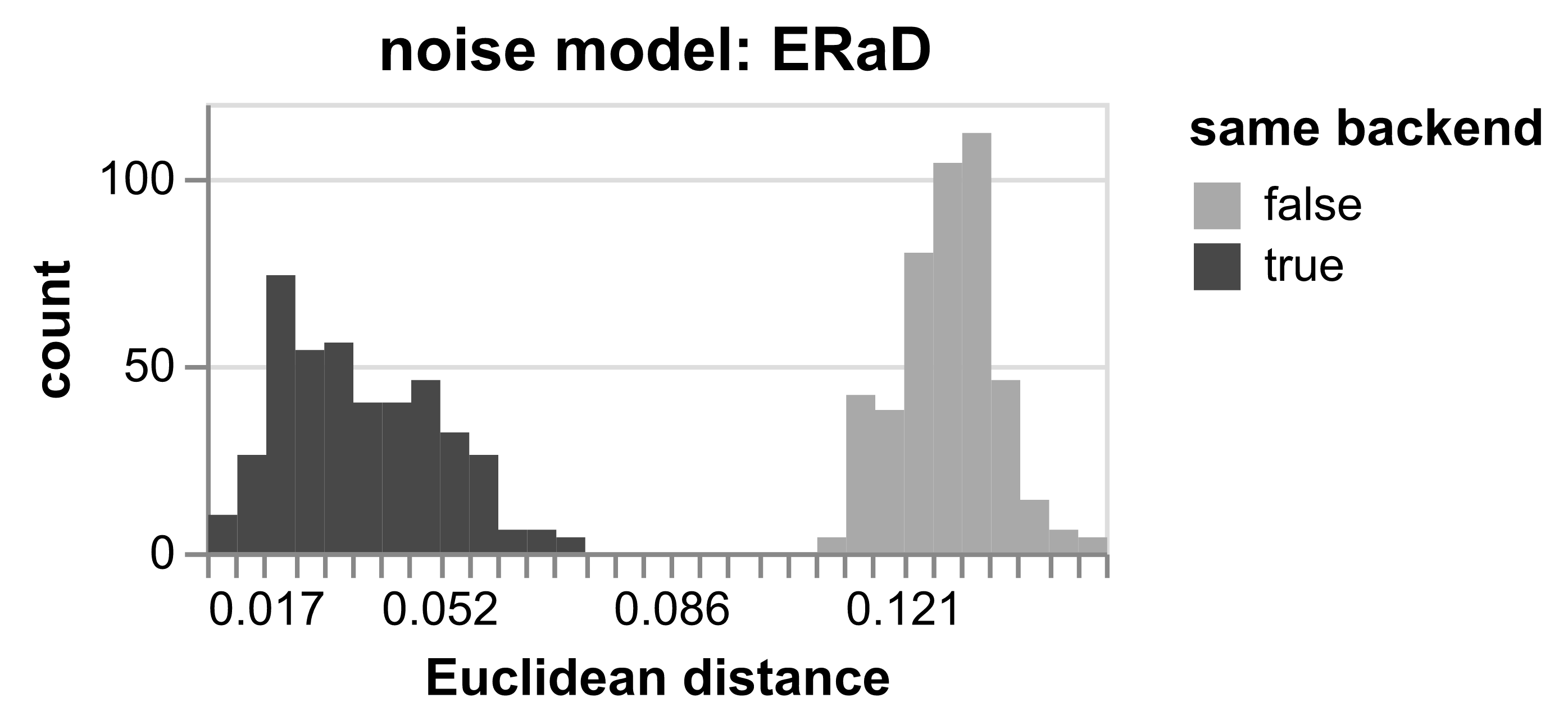}
    \caption{Shown above is a comparison between shots simulated on \texttt{ibm\_brisbane} and \texttt{ibm\_sherbrooke} under the ERaD noise model. The plot shows distributions of Euclidean distances between pairs of jobs (all shots averaged down to one feature vector), with the pairs divided into being from either the same or from different backends. We see more separation between same versus different backends than in \cref{fig:euclidean-complex}.}
    \label{fig:euclidean-thermal}
\end{figure}

Counterintuitively, when we test certain pairs of backends individually, we find that the ERaD model yields much higher distinguishability than full backend emulation. For example, we compare simulated fingerprints from \texttt{ibm\_brisbane} and \texttt{ibm\_sherbrooke}. We run roughly 8,000 shots per backend per noise model, split across 14 jobs per backend. For each job, we average the 512 output bit vectors element-wise into one feature vector. We then compare it pairwise with every other representative feature vector and find the Euclidean distance between each pair. Histograms of these distances are plotted in \cref{fig:euclidean-thermal,fig:euclidean-complex}.

Clearly, we see greater separation between same-backend and different-backend pairs in the ERaD model, despite lower overall classification accuracy as shown in \cref{fig:simulation}. We speculate that the more sophisticated backend-emulation model may incorporate an error channel that has greater dynamic range than the channels in the ERaD model; this could account for ERaD providing better distinguishability between two backends, while backend emulation provides better distinguishability between five backends.

The emulation of these errors does not completely capture the classification performance of real hardware. Both ERaD and backend emulation utilize a Markovian approach to noise emulation which assumes weak coupling between the quantum system and the environment. Markovian models perform demonstrably well for simulating hardware in some cases \cite{georgopoulos2021modeling}, but real quantum devices are partially governed by non-Markovian dynamics \cite{de2017dynamics} which necessitates different noise models \cite{agarwal2024modelling} for more complete simulations. Investigating noise models which account for non-Markovianity as emulators of classification performance may provide greater understanding of the information channels leveraged for fingerprinting in this work.

\section{Security Analysis and Limitations}

\label{sec:limitations}
We consider the strategies that an attacker may use to cause our protocol to erroneously authenticate a fingerprint. Because we assume that providers have cleartext access to user circuits, under the Overloaded Provider model, the provider may attempt to obfuscate error information by transforming the submitted circuit into a semantically equivalent quantum circuit with different error properties than the original. One way to accomplish this is to randomly transform the circuit by locally replacing gates with different, but equivalent gates. While this is computationally straightforward, it will likely degrade the accuracy of the circuit (since it was likely already optimized).

Another option for a malicious provider (either first-party or third-party reseller) is to implement a transparent layer of error correction, so that the qubits offered to users are not physical, but logical qubits. Quantum computers will likely have to use several of these layers of error correction, where each layer forms more-reliable logical qubits out of noisier qubits offered by the layer beneath it, and physical qubits form the bottom layer. Implementing a couple layers of error correction may decrease the fingerprint signal in any error syndrome measurements that users submit in their own circuits.

Either of these methods should at least be detectable to the user. If the provider randomizes the circuit, the error fingerprint will change on each execution. If the provider implements a transparent layer of error correction, the error magnitude will be significantly lower than expected.

Under the Dishonest Reseller model, neither of the circumventions we mentioned above apply because we assume that the first-party provider cooperates with the user in providing expected fingerprints for their hardware. Any divergence from these expected fingerprints is then considered a verification failure, so attackers would have to resort to modeling attacks on the PUF scheme itself.

\paragraph{Modeling Attacks} The security of PUF constructions depends on the prover's inability to ``learn'' the fingerprinting function, i.e.~the function must be \emph{unclonable}. In our construction, learning the fingerprinting function amounts to being able to predict the distribution of error syndrome measurements given an input circuit. Intuition suggests this may be difficult because it is equivalent to classical simulation of a quantum circuit, which is believed to be exponentially hard in the number of qubits \cite{feynman1982sim,harrow2017supremacy}. However, not all quantum circuits are hard to simulate. Circuits comprised of only Clifford gates can be simulated in polynomial time ($O(n^2)$ in the number of qubits) \cite{gottesman1998clifford}. Pirnay~et~al.~\cite{pirnay2022learning} additionally demonstrate a modeling attack on quantum fingerprints that are determined by the classical readout of circuits using only single-qubit gates (even including non-Clifford gates), so simply relying on the ``quantumness'' of outputs is insufficient.

There is reason to believe our proposed construction is difficult to learn. Our QPUF relies partially on non-Markovian dynamics, which are known to be difficult to simulate \cite{rivas2014markovian, devega2017markovian}. If a system obeys non-Markovian dynamics, then its evolution depends not only on the current state but also on its full history, which is intrinsically tied to the correlations built up between the system and its environment. In the quantum setting these correlations are complicated by the entanglement between system and environment, making it difficult to isolate a clean dynamical map for the system alone. Capturing these effects requires complex models \cite{rivas2014markovian}. How to accurately and efficiently simulate non-Markovian dynamics in quantum systems is an open problem; de~Vega~et~al.~\cite{devega2017markovian} survey the field as of 2017.

In theory, our fingerprint construction incorporates non-Markovian dynamics because it characterizes all errors that occur in the circuit (within the limit of the chosen error correcting code, which users will choose to be able to reliably detect and correct all errors). As long as non-Markovian errors remain a significant source of error \cite{breuer2016markovian} in quantum systems, our method will incorporate that information in error syndromes and be resistant to modeling attacks.

\section{Related Work}
\label{sec:related-work}

\subsection{Quantum PUF}

There have been several proposed fingerprinting methods claiming to be able to identify quantum hardware \cite{mi2021short, phalak2021puf, smith2023fast, bathalapalli2023qpuf}. Similarly, there have been proposed constructions for physical unclonable functions. To the best of our knowledge, all prior QPUF constructions require quantum communication channels between prover and verifier; use expensive, dedicated fingerprinting circuitry; or are vulnerable to modeling attacks.

Mi~et~al.~\cite{mi2021short} propose using idle tomography (IDT) to generate a fingerprint of a quantum device. IDT is a diagnostic technique in quantum computing to characterize noise and errors that affect quantum systems when they are not actively performing operations \cite{blume2019idle}. Smith~et~al.~\cite{smith2023fast} analyze several methods of fingerprinting, and propose that taking the set of qubit resonant frequencies of a quantum device at a given time is an effective fingerprint. Both of the designs put forward by Mi~et~al.~\cite{mi2021short} and Smith~et~al.~\cite{smith2023fast} are expensive to run, and require running a dedicated fingerprinting circuit that is not tied to an otherwise meaningful computation. This incurs large cost overhead and opens the door to adversarial providers trying to predict which circuits are for fingerprinting so they can route only those to the correct backends.

Phalak~et~al.~\cite{phalak2021puf} propose two QPUF architectures --- one based on superposition bias and another on decoherence decay --- to authenticate quantum hardware. They demonstrate that error rates, decoherence times, and readout biases vary sufficiently across devices to serve as stable hardware fingerprints. They use separate, dedicated circuits to calculate their fingerprints, which are expensive and may be detected by a malicious provider and handled honestly while dishonestly executing business circuits on inferior hardware (what we call a routing attack).

Bathalapalli~et~al.~\cite{bathalapalli2023qpuf} propose a quantum QPUF architecture designed for Security-by-Design in Industrial Internet-of-Things (IIoT) environments. Their implementation uses Hadamard and $R_y$ gates on IBM quantum devices to generate device-specific identities, and incorporates a majority-vote scheme to reduce noise. The architecture supports a large challenge-response space, but as discussed next is based on single-qubit gates, so is learnable as shown by Pirnay~et~al.~\cite{pirnay2022learning}.

An important restriction on the design space was put forward by Pirnay~et~al.~\cite{pirnay2022learning}. They demonstrate an attack on classical-readout quantum physical unclonable functions (CR-QPUF) based on single-qubit gates, such as the $H$ and $R_y$ gates used by Bathalapalli~et~al.~\cite{bathalapalli2023qpuf}. They show that an adversary constrained to the statistical query model \cite{kearns1998sq} (having only oracle access to probability estimates of quantum circuit outcomes) can forge the signature of a quantum device through a modeling attack using simple regression of low-degree polynomials. This attack allows the adversary to model and predict the CR-QPUF characteristics, effectively forging the quantum device's fingerprint.

Goorden~et~al.~\cite{goorden2014qreadout} introduce an authentication protocol using physical unclonable keys composed of multiple-scattering optical media. By verifying the complex speckle-like response to a challenge consisting of optical input, the scheme exploits quantum limitations on measurement to prevent digital emulation attacks, even when an adversary knows the full challenge–response database. The approach is secure without relying on secret data or computational assumptions and can be implemented with existing photonic technology. However, it relies on having an optical communication channel between the prover and verifier.

\subsection{Classical PUF}

Pappu~et~al.~\cite{pappu2002puf} (2002) introduce the concept of physical one-way functions, leveraging the unique scattering properties of disordered optical media to create unclonable identifiers. Unlike traditional cryptographic one-way functions, which rely on unproven conjectures, the authors' construction exploits the inherent complexity of light scattering in mesoscopic structures to generate unique, tamper-resistant responses to optical challenges. The authors demonstrate an authentication system where a laser probe interrogates a token containing randomly distributed scatterers, producing a speckle pattern that serves as a secure identifier. Their approach provides a physically unclonable security primitive with a large address space, making duplication computationally infeasible. The work is foundational for physical unclonable functions (PUFs), influencing subsequent research in hardware security and device authentication.

In the same year, Gassend~et~al.~\cite{gassend2002silicon} introduce the concept of silicon-based PUFs as a method for uniquely identifying and authenticating integrated circuits. Instead of optical techniques, their approach leverages inherent manufacturing process variations in silicon to generate unique challenge-response pairs. The authors demonstrate the feasibility of this method using FPGA implementations and propose techniques to mitigate environmental variations and adversarial modeling attacks. Their work establishes the foundation for silicon-based PUFs as a robust mechanism for secure authentication, smartcard security, and anti-counterfeiting applications.

Guajardo~et~al.~\cite{guajardo2007ip} introduce the concept of ``strong'' versus ``weak'' PUFs, referring to PUFs with a large and small challenge space, respectively. The authors leverage the inherent randomness of SRAM startup values to create a strong PUF that uniquely identifies each FPGA without requiring additional hardware modifications. This PUF is then used to derive a cryptographic key, which can be reliably reconstructed using a fuzzy extractor to correct for measurement noise. This key is either shared with a trusted third party (TTP) or used locally to encrypt and authenticate FPGA bitstreams. By binding the IP to the unique PUF response of a given FPGA, the system prevents cloning attacks, ensures hardware authentication, and maintains confidentiality and integrity of the IP, thereby securing FPGA-based designs from unauthorized duplication or tampering.

\section{Conclusions}
\label{sec:conclusion}

In this work, we introduce a framework for leveraging error syndrome measurements --- already integral to quantum error correction~---~as a tool for quantum device fingerprinting and authentication. Using the output of syndrome measurements, we effectively embed verification within standard quantum computations, eliminating the need for expensive, specialized verification circuits. Our results on multiple IBM quantum backends --- over varied starting states, optimization levels, qubit mappings, and error-correcting codes (Shor, Steane, and the surface code) --- demonstrate that syndrome-based classifiers can consistently identify between five backends with over 99\% accuracy using fewer than 500 shots for inference. Fingerprinting through error syndromes is effective, inexpensive, and integrates directly into the quantum computation pipeline. Our framework will encourage further research into the design of robust and convenient security primitives that will serve as a foundation of trust in quantum systems.

\bibliographystyle{ACM-Reference-Format}
\bibliography{citations}

\end{document}